\documentclass[aps,preprint,showkeys]{revtex4-1}
\usepackage{graphicx}
\usepackage{amsmath}
\usepackage{makeidx}
\usepackage{amsfonts}
\usepackage{amssymb}
\usepackage{mathtools}
\usepackage{xcolor}

\begin{document}

\title{Noise tolerance via reinforcement in the quantum search problem}
\date{\today}

\author{Marjan Homayouni-Sangari}
\email{marjan.homayouni.sangari@gmail.com}
\affiliation{Department of Physics, College of Science, Shiraz University, Shiraz 71454, Iran}

\author{Abolfazl Ramezanpour}
\email{aramezanpour@gmail.com}
\affiliation{Department of Physics, College of Science, Shiraz University, Shiraz 71454, Iran}
\affiliation{Leiden Academic Centre for Drug Research, Faculty of Mathematics and Natural Sciences, Leiden University, PO Box 9500-2300 RA Leiden, The Netherlands}

\begin{abstract}
The Grover lower bound for the unstructured search problem can be surpassed when some information about the data structure is available. Here, we numerically observe that reinforcement can exponentially reduce the number of required evolution layers from $\sqrt{D}$ to $\ln D$ in a $D$-dimensional system, by exploiting the information provided by the quantum state. Therefore, a reinforced quantum search is expected to exhibit a larger noise threshold compared to a standard search algorithm in a noisy environment. We use numerical simulations to characterize the level of noise tolerance via reinforcement in the presence of both coherent and incoherent noise, considering a system of $N$ qubits and a single $D$-level (qudit) system. Our results show that reinforcement significantly enhances the algorithm's success probability and improves the scaling of the number of reinforced evolution layers with system size. These findings indicate that reinforcement offers a promising strategy for error mitigation, especially when a precise noise model is unavailable.
\end{abstract}

\maketitle

\section{Introduction}\label{S0}
Quantum computations are inherently sensitive to various sources of noise, including environmental interactions, control imperfections, and crosstalk between qubits. Understanding how quantum algorithms perform under such realistic, non-ideal conditions is therefore crucial for assessing their theoretically predicted speedups in practice. This issue is particularly pressing in the current noisy intermediate-scale quantum (NISQ) era, where qubit coherence times are limited, gate fidelities are imperfect, and full fault-tolerant error correction remains experimentally challenging~\cite{Knill-nat-2005,Preskill2018,Bharti2022,Kim-nat-2023}.

Incoherent noise, such as dephasing, depolarization and amplitude damping, affects the quantum state in a random and probabilistic manner, gradually reducing the state fidelity over time~\cite{Sun-pap-2021,Urbanek-prl-2021,P-acm-2025}. Coherent noise originates from systematic errors such as over-rotations, calibration drifts, crosstalk and other unitary imperfections~\cite{Green-qst-2017,Cai-qi-2020}. Both types of noise can have significant impacts on the performance of quantum algorithms. For instance, in Grover's search algorithm, noise can substantially reduce the success probability and, in certain regimes, eliminate the quadratic speedup that underlies the algorithm’s advantage~\cite{Pablo-pra-1999,Shenvi2003,Shapira2003,Regev2008,Salas2008}. In variational quantum algorithms, noise can lead to barren plateaus in the optimization landscape, hindering the ability to find optimal solutions~\cite{McClean2018,Wang2021}.

To address these challenges, a variety of error mitigation techniques have been developed, particularly for near-term quantum devices \cite{QEM-rmp-2023,Aharonov-arxiv-2025}. These methods aim to reduce the impact of noise without requiring full-scale quantum error correction. Techniques such as zero-noise extrapolation, probabilistic error cancellation, and machine learning have been successfully applied to improve the performance of quantum algorithms in noisy settings~\cite{Geller2013,Temme2017,Li-prx-2017,Endo2021,Van-np-2023,Liao-nml-2024}. These challenges also highlight the need for adaptive strategies that can respond to complex or unknown noise~\cite{Fosel2018,Lin-pra-2020,Strikis-prx-2021}. In this paper, we show that reinforcement (a feedback-based strategy) can significantly improve the success probability of a quantum search algorithm in the presence of coherent/incoherent noise.

Reinforcement algorithms have emerged as a powerful paradigm in the study of classical optimization problems ~\cite{SB-book-1998,Alf-prl-2006,Mazyavkina2021}. By amplifying actions that lead to favorable outcomes and suppressing those that do not, reinforcement can accelerate convergence toward optimal solutions and improve robustness against noise and errors ~\cite{QR-pra-2017,QR-jstat-2025}. Reinforcement approaches have been successfully incorporated into various problems such as parameter optimization, circuit simplification, preparation of complex quantum states, and improved exploration of rugged energy landscapes~\cite{Fosel2018,Bukov2018,Ayanzadeh2020,Fosel2021,QR-pra-2022}. These developments highlight the versatility of reinforcement in classical and quantum optimization problems and suggest their potential utility for mitigating noise and enhancing the practical performance of quantum algorithms.

To be specific, we shall consider the (unstructured) quantum search problem, where Grover's algorithm provides the optimal computation time in the absence of noise ~\cite{grover-prl-1997,Roland-pra-2002,fdsearch-prl-2005,fdsearch-pra-2017,search-pra-2020}. In this ideal case, Grover's procedure amplifies the probability of finding a marked state through a sequence of rotations within a two-dimensional subspace, reaching its maximum success probability after approximately $\sqrt{D}$ iterations, where $D$ denotes the dimension of the search space. In presence of noise, these precise rotations are perturbed, leading to a reduced peak success probability and a shift in the optimal measurement time~\cite{Pablo-pra-1999,Shenvi2003,Shapira2003,Regev2008,Salas2008,Pan-qi-2023,Pati-jpa-2024,Leng-prr-2025}. For instance, systematic imperfections in the Hadamard gates and the diffusion operator have been shown to rapidly degrade performance unless the error amplitude is less than $O(D^{-1/4})$ ~\cite{Shapira2003}.

In this study, we introduce a reinforcement strategy which is aimed at mitigating errors in the quantum search problem. The main idea is to use the current quantum state of the system as feedback in the Hamiltonian, thereby energetically favoring the trajectory followed by the unperturbed dynamics. This same mechanism also mitigates the detrimental effects of quantum transitions to excited states in the quantum annealing algorithm ~\cite{QR-pra-2022}. Unlike most conventional error-mitigation techniques, our approach does not require a detailed characterization of the noise model a priori. This makes it particularly well suited for experimental platforms where noise is complex, time-dependent, or difficult to model accurately. The main challenge is
therefore the accurate and efficient inference of the quantum state during the reinforced evolution. While an exact treatment of such quantum feedback is a formidable task, approximate reinforcement algorithms can be implemented using techniques developed for collective weak measurements and quantum principal component analysis \cite{Haah-acm-2016,QFC-pra-1999,lloyd-pra-2000,EQD-pra-2001,QFC-prl-2010,lloyd-nphys-2014}. In what follows, we assume that the quantum state is available at each stage of the search algorithm.

The paper is organized as follows. Section~\ref{S1} gives the main definitions and the problem statement. We then investigate the effects of reinforcement on the quantum search problem in ideal (noise-free) and noisy settings. Section~\ref{S2} considers the ideal case, while Secs.~\ref{S3} and ~\ref{S4} focus on coherent and incoherent noise. In both noisy scenarios, we study two representative examples: one based on a system of $N$ qubits and the other involving a single $D$-level system (a qudit). The concluding remarks are given in Sec.~\ref{S5}.

\begin{figure}
\includegraphics[width=11cm]{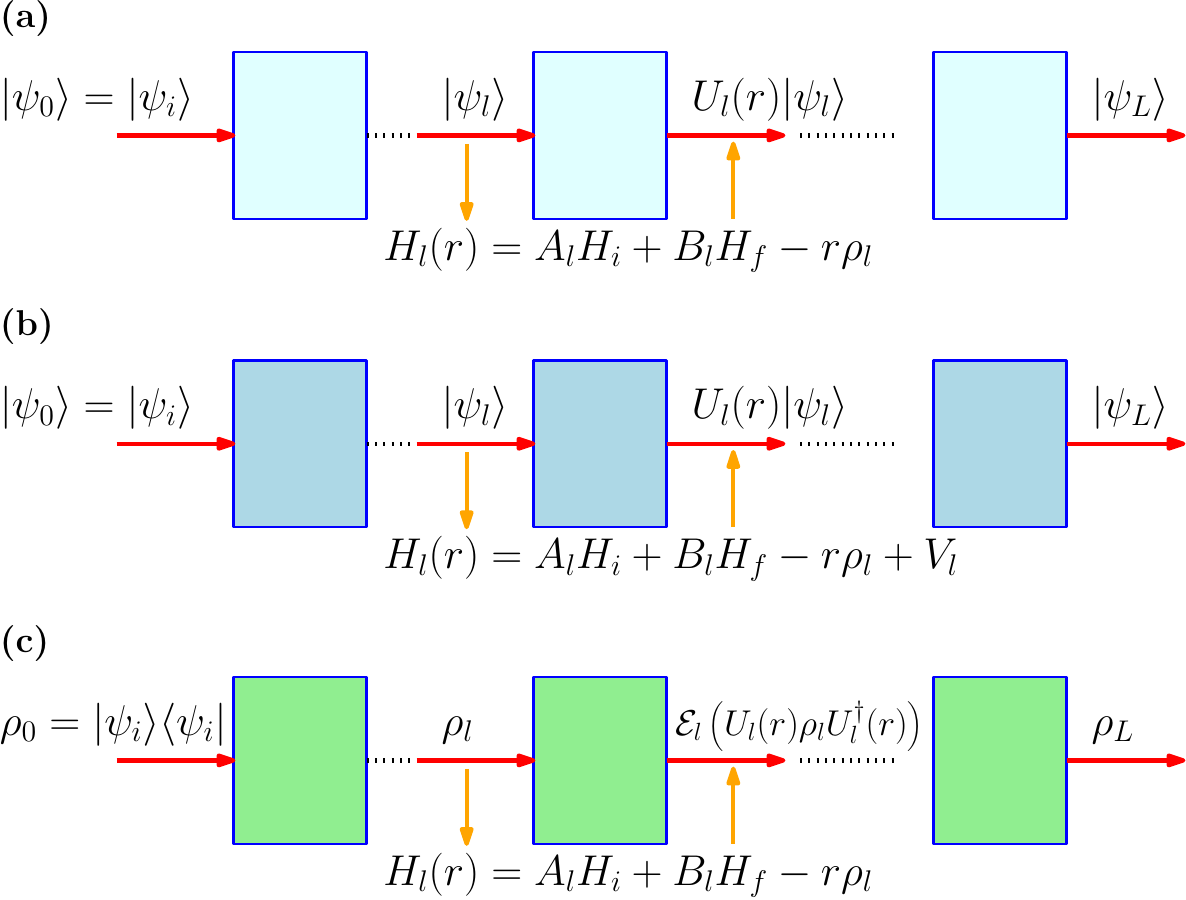} 
\caption{An illustration of $L$ layers of quantum evolution with reinforced Hamiltonians $H_l(r)$. (a) Coherent evolution in the absence of noise. Here $\rho_l=|\psi_l\rangle\langle \psi_l|$ and $U_l(r)=e^{-\mathrm{i}H_l(r)}$. (b) Coherent evolution in the presence of noise $V_l$. (c) Evolution in the presence of incoherent noise, represented by channels $\mathcal{E}_l$.}\label{fig0}
\end{figure}

\section{Problem Statement}\label{S1}
Consider a quantum system of $N$ qubits in a Hilbert space of dimension $D=2^N$ with basis states $|\boldsymbol\sigma\rangle$. Here  $\boldsymbol\sigma=\{\sigma_i=0,1,i=1\dots,N\}$ and the $|\sigma_i\rangle$ are eigenstates of Pauli operators $\sigma_i^{z}$.
The quantum search problem can be formulated as a quantum annealing (QA) process from the ground state
of $H_i=\mathbb{I}-|\psi_i\rangle\langle \psi_i|$ to that of $H_f=\mathbb{I}-|\psi_f\rangle\langle \psi_f|$ \cite{Roland-pra-2002,fdsearch-pra-2017}. Here $|\psi_f\rangle$ is the target state and 
\begin{align}
	|\psi_i\rangle=\frac{1}{\sqrt{2^N}}\sum_{\boldsymbol\sigma}|\boldsymbol\sigma\rangle=\sqrt{P_0}|\psi_f\rangle+\sqrt{1-P_0}|\psi_f^{\perp}\rangle.
\end{align}
We use $|\psi_f^{\perp}\rangle$ to represent the subspace of states that are orthogonal to $|\psi_f\rangle$.
$P_0=1/2^N$ is the probability of finding the target state in $|\psi_i\rangle$.

The annealing process is approximated by $L$ layers of quantum evolutions, see Fig.\ref{fig0}. The Hamiltonian in each layer $l=0,\dots,L-1$ is
\begin{align}
	H_l(r)=A_lH_i+B_lH_f+R_l(\rho_l)+V_l,
\end{align}
with the reinforcement term $R_l(\rho_l)$ and (possibly) the coherent noise $V_l$. The reinforcement term is an increasing function of the quantum state of the system $\rho_l$ at layer $l$. A reasonable example is $R_l(\rho_l)=-r_l\ln(\rho_l)$ \cite{QR-jstat-2025}. By assigning lower energies to the more probable states, this reinforcement is expected to increase the minimal energy gap during the annealing process. Consequently, it reduces the likelihood of quantum transitions from the instantaneous ground state to excited states. Moreover, it can even alter the nature of the quantum phase transition, transforming it from a discontinuous one characterized by an exponentially small energy gap into a continuous one with a polynomially small gap in system size \cite{QR-pra-2017,QR-pra-2022}.

For simplicity, in the following we work with $R_l(\rho_l)=-r_l\rho_l$. Note that for pure states, any function of $\rho_l=|\psi_l\rangle\langle \psi_l|$ that has a Taylor expansion can be written as a linear function of $\rho_l$. The reinforcement term $-r_l\rho_l$, with $r_l>0$, reduces the expected energy of unperturbed states that have a high overlap with the current system state. It should be noted, however, that very large reinforcement values ($r_l\gg 1$) can freeze the system state and undermine the advantages of the annealing algorithm. Consequently, there exists an optimal range for the reinforcement parameter within which the algorithm performs most effectively.

The dynamics starts with the ground state of $H_i$, that is $\rho_0=|\psi_i\rangle\langle \psi_i|$. In each layer, we use coherent/incoherent evolutions (represented by unitary $U_l(r)=e^{-\mathrm{i}H_l(r)}$ and quantum map $\mathcal{E}_l$) to obtain $\rho_{l+1}=\mathcal{E}_l(U_{l}\rho_l U_{l}^{\dagger})$ from $\rho_l$. The success probability at each layer is obtained from $P_{success}(l)=\mathrm{tr}(\rho_{l}|\psi_f \rangle\langle \psi_f|)$. The question is how much the reinforcement term can enhance the success probability in presence of coherent or incoherent noise.  

For an annealing process, the Hamiltonian coefficients $A_l=1-t_l$ and $B_l=t_l$, where $t_l\in (0,1)$ is a monotonically increasing function of layer index $l$. The optimal annealing schedule, in the absence of noise and reinforcement, is provided by the Grover algorithm \cite{Roland-pra-2002,fdsearch-pra-2017}
\begin{align}\label{grover}
	t_l^G &=\frac{1}{2}\left[1-\sqrt{\frac{P_0}{1-P_0}}\tan\left((1-2\frac{l}{L-1})\phi\right)\right],\\
	\phi &=\arctan\left(\sqrt{\frac{1-P_0}{P_0}}\right).
\end{align}
In this case the Hamiltonian coefficients are denoted by $A_l^G, B_l^G$.
Besides the annealing algorithm, we also study a locally optimal process where at each layer $l$ the Hamiltonian coefficients are optimized to maximize the success probability at that layer. The optimal coefficients in this case are denoted by $A_l^*, B_l^*$.

\section{Reinforced dynamics in absence of noise}\label{S2}
In the absence of noise the dynamics is governed by
\begin{align}
	H_l(r)=A_lH_i+B_lH_f-r_l\rho_l,
\end{align}
where $\rho_l=|\psi_l\rangle\langle \psi_l|$ (Fig.\ref{fig0} panel (a)). 
The quantum states $|\psi_l\rangle$ evolve by $|\psi_{l+1}\rangle=U_l(r)|\psi_{l}\rangle$ starting from $|\psi_0\rangle=\sqrt{P_0}|\psi_f\rangle+\sqrt{1-P_0}|\psi_f^{\perp}\rangle$.
Here the system state is limited to a two-dimensional subspace which is spanned by $\{|\psi_f\rangle,|\psi_f^{\perp}\rangle\}$.
Let us write the quantum state as $|\psi_l\rangle =\alpha_l|\psi_f\rangle+\beta_l|\psi_f^{\perp}\rangle$,
with Pauli operators acting on the two-dimensional subspace.
The unitary operator at layer $l$ is 
\begin{align}
	U_l(r) =e^{-\mathrm{i}H_l(r)}=e^{-\mathrm{i}h_{l0}}\left(\cos(h_l)\mathbb{I}-\mathrm{i}\sin(h_l)\hat{h_l}.\vec{\sigma}\right),
\end{align}
where the field components are given by
\begin{align}
	h_{l0}(r_l) &=\frac{1}{2} \left(A_l+B_l -r_l\right) ,\\
	h_{lx}(r_l) &=-A_l\sqrt{P_0(1-P_0)}-r_l\operatorname{Re}(\alpha_l\beta^*_l),\\
	h_{ly}(r_l) &=r_l\operatorname{Im}(\alpha_l\beta^*_l),\\
	h_{lz}(r_l) &=\frac{1}{2}[A_l-B_l-2A_lP_0-r_l(2|\alpha_l|^2-1)],
\end{align}
and $h_l =\sqrt{h_{lx}^2+h_{ly}^2+h_{lz}^2}$. We see how the current state of the system affects the magnitude and direction of rotations in the two-dimensional space of the problem.

\begin{figure}
\includegraphics[width=14cm]{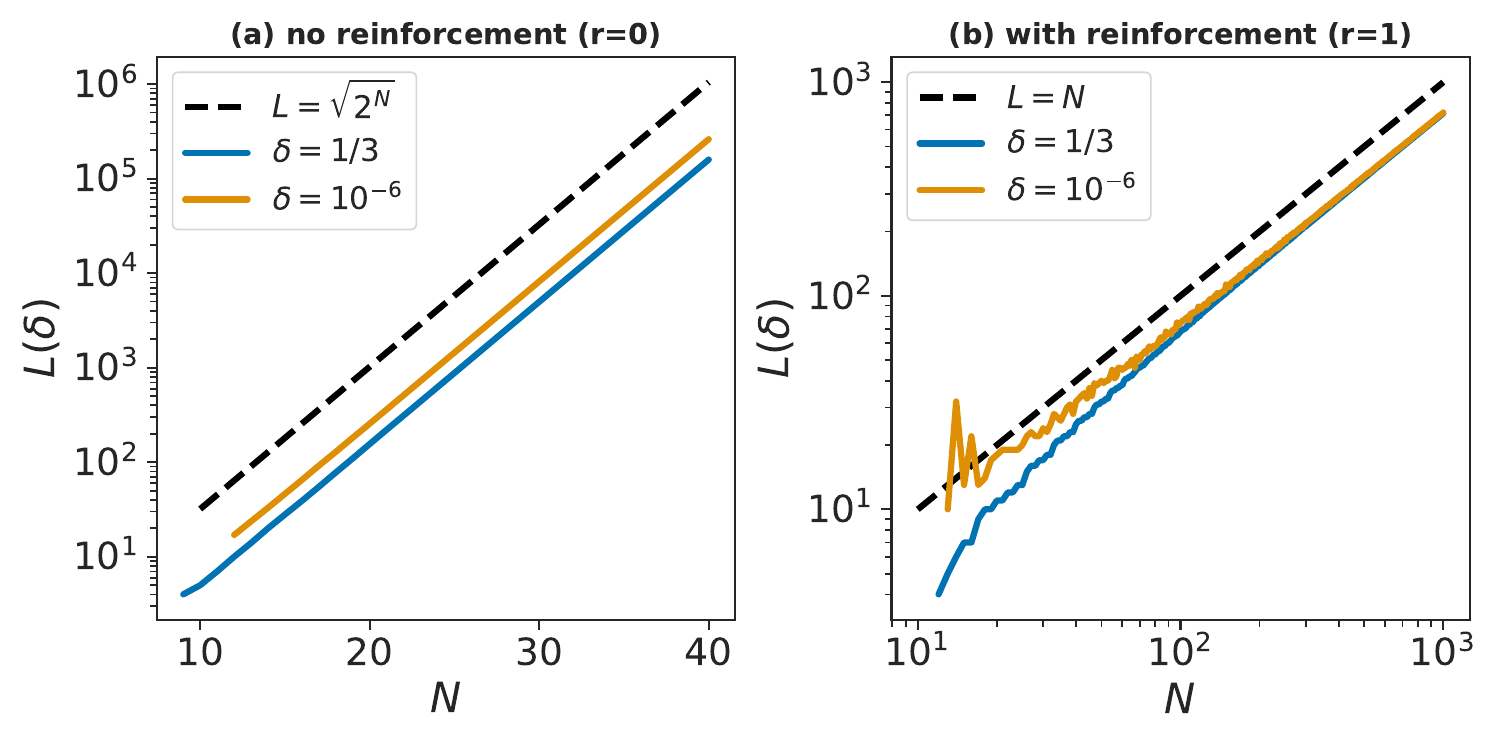} 
\caption{Quantum search in the absence of noise: The required number of evolution layers $L(\delta)$ vs the system size $N$ (a) without reinforcement $r=0$, (b) with reinforcement $r=1$. $L(\delta)$ is the minimum number of evolution layers needed to have $P_{success}>1-\delta$.}\label{fig1}
\end{figure}

In each layer, we obtain the optimal parameters $A_l^*, B_l^*$ by maximizing the success probability at layer $l+1$. This is accomplished via an exhaustive search over $K^2$ discrete parameter values, subject to the constraint $0\le A_l,B_l\le 2\pi$. The time complexity of this local optimization is $LK^2$ for $L$ layers. Here, we set $K=10$. For simplicity, we also take $r_l=r$, independent of the layer index $l$.
Figure \ref{fig1} shows the minimum number of layers $L(\delta)$ needed to have a success probability that is greater than $1-\delta$. Without reinforcement ($r=0$) we recover the Grover scaling for the computation time, that is $L(\delta)\propto \sqrt{2^N}$ as expected. As panel (b) shows, with reinforcement ($r=1$) the required number of layers reduces to $L(\delta)\propto N$ for different values of $\delta=1/3,10^{-6}$. This is the main numerical observation of the paper: within the state-dependent feedback model, reinforcement can exponentially reduce the number of evolution layers. Note that it is not a claim of reduced oracle/query complexity in the standard circuit model. In the following sections we shall see how reinforcement affects the performance of a quantum search problem in the presence of coherent and incoherent noise.

\section{Reinforced dynamics of an N-qubit system }\label{S3}
Consider a system of $N$ qubits and the computational basis $|\boldsymbol\sigma\rangle=|\sigma_1,\dots,\sigma_N\rangle$.
Let us represent the target state by $|\psi_f\rangle =|\mathbf{0}\rangle$, so $|\psi_f^{\perp}\rangle =\frac{1}{\sqrt{2^N-1}}\sum_{\boldsymbol\sigma\ne \mathbf{0}}|\boldsymbol\sigma\rangle$. The initial state is $|\psi_i\rangle=\sqrt{P_0}|\psi_f\rangle+\sqrt{1-P_0}|\psi_f^{\perp}\rangle$ with $P_0=1/2^N$.

\subsection{Coherent noise}\label{S31} 
At each layer $l$ we construct the reinforced Hamiltonian $H_l(r)=A_lH_i+B_lH_f-r_l\rho_l+V_l$ with quantum state $\rho_l =|\psi_l\rangle\langle \psi_l|$ (Fig.\ref{fig0} panel (b)).  For coherent noise $V_l$, we simply take the Pauli operators of weight 1:
\begin{align}
	V_l=\sum_{i=1}^N\sum_{\mu=x,y,z}\epsilon_{l,i\mu}\sigma_i^{\mu}.
\end{align} 
The coefficients $\epsilon_{l,i\mu}=\mathcal{N}(0,\epsilon_l)$ are independent random variables drawn from a normal distribution of mean zero and variance $\epsilon_l^2$. We work with the scaling $\epsilon_{l}=\epsilon/L$.
Thus $L$ is used here as the number of discretization layers for a fixed integrated noise strength $\epsilon$, rather than as a fixed noise rate per physical layer. The quantum states evolve by $|\psi_{l+1}\rangle=U_l(r)|\psi_{l}\rangle$ starting from $|\psi_i\rangle$.

\begin{figure}
	\includegraphics[width=12cm]{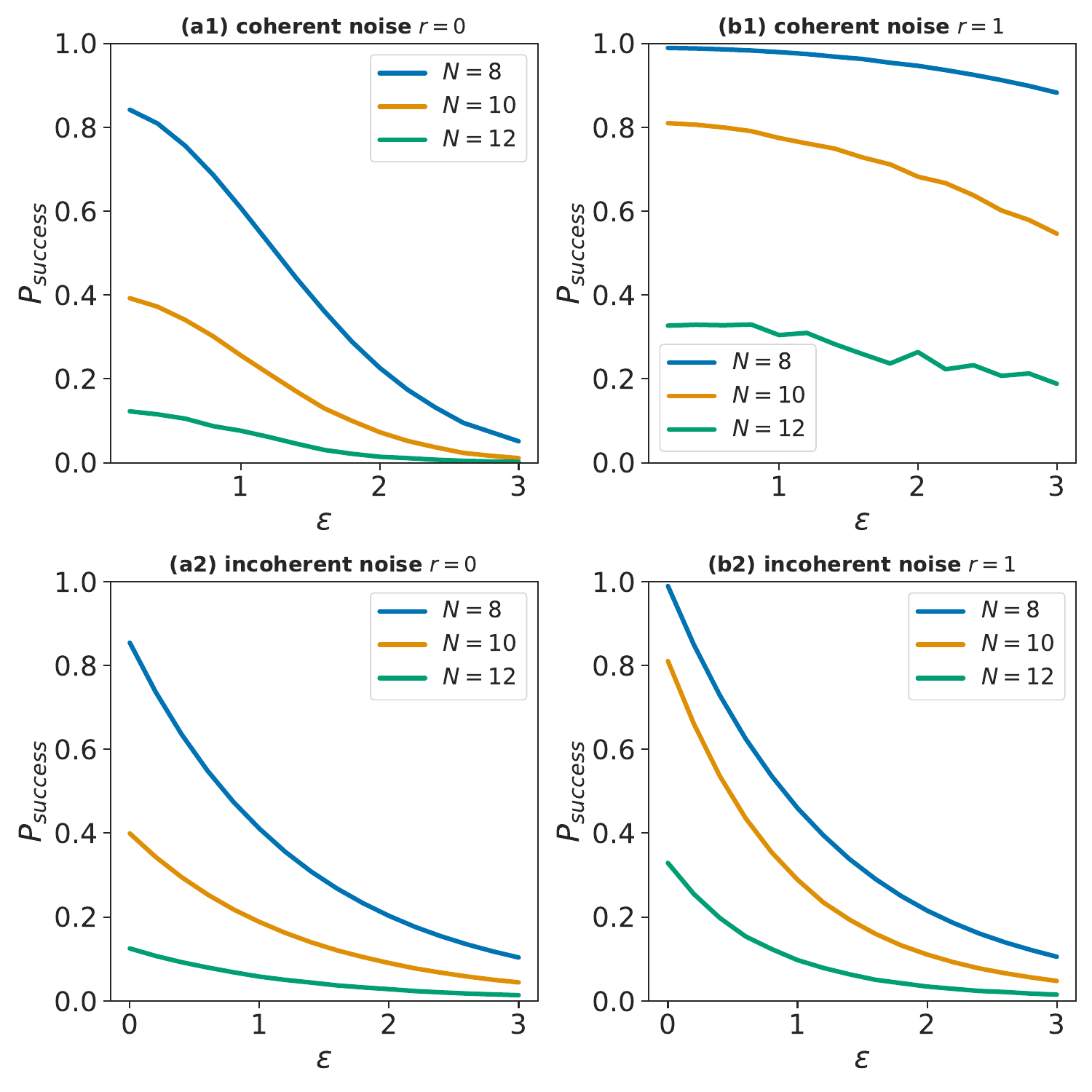} 
	\caption{Quantum annealing with coherent and incoherent noise in a system of $N$ qubits using the Grover coefficients $A_l^G,B_l^G$: Success probability $P_{success}(L)$ vs the noise strength $\epsilon$ for $r=0$ ((a1),(a2)) and $r=1$ ((b1),(b2)). Here $L=50$. The points are results of averaging over at least $20$ independent realizations of the noisy Hamiltonian. Statistical errors are less than $0.05$ for coherent noise and $10^{-3}$ for incoherent noise.}\label{fig2}
\end{figure}

The success probability is computed for different noise realizations to obtain the average success probability in a quantum annealing process using the Grover coefficients. Figure \ref{fig2} (upper panels) shows $P_{success}(L)$ at the end of the annealing process vs $\epsilon$ for different system sizes $N=8,10,12$, in order to compare the degradation of the success probability with and without reinforcement. 
We see that for $N=12$ the success probability of the reinforced dynamics ($r=1$) is at least two times greater than that of the standard dynamics ($r=0$). Interestingly, the effect of reinforcement on the performance is more pronounced at higher values of noise strengths $\epsilon \in (0,3)$.  

\subsection{Incoherent noise}\label{S32}
Here we consider $L$ consecutive layers of unitary evolutions $U_l$ and quantum maps $\mathcal{E}_l$ for $l=0\cdots,L-1$ (Fig.\ref{fig0} panel (c)). The maps $\mathcal{E}_l$ are to model noisy interactions with an environment. For an input state $\rho_0$, we obtain the sequence of states $\rho_{l+1}=\mathcal{E}_l(U_{l}\rho_l U_{l}^{\dagger})$ up to the output state $\rho_{L}$. 

As before, the unitary part of the evolution $U_l=e^{-\mathrm{i}H_l}$ represents a reinforced quantum dynamics. That is $H_l =A_lH_i+B_lH_f-r\rho_l$ with ground states $|\psi_i \rangle$ and $|\psi_f \rangle$ for the initial and final Hamiltonians. More precisely, the quantum state here evolves as follows
\begin{align}
\rho_{l+1} =(1-\epsilon_l)U_l\rho_l U_l^{\dagger}+\epsilon_l\sum_{i=1}^N\sum_{\mu=x,y,z}w_{i\mu}\sigma_i^{\mu}(U_{l}\rho_l U_{l}^{\dagger})\sigma_i^{\mu},
\end{align}
starting from $\rho_0=|\psi_i \rangle\langle \psi_i|$. 
The parameters $\epsilon_l\in (0,1)$ and iid random variables $\{w_{i\mu}\in (0,1) :\sum_{i\mu} w_{i\mu}=1\}$ control the strength and structure of Pauli noise. We set $\epsilon_l=\epsilon/L$.

We compute the average of success probability $P_{success}(L)=\mathrm{tr}(\rho_{L}|\psi_f \rangle\langle \psi_f|)$ at the end of a quantum annealing process using the Grover coefficients $A_l^G$ and $B_l^G$. Figure \ref{fig2} (lower panels) shows the success probability for different noise strengths and number of qubits $N=8,10,12$. We observe that incoherent noise is more detrimental than coherent noise in the quantum search problem. Nevertheless, reinforcement significantly enhances the algorithm’s performance, with the effect being more pronounced in this case at lower noise strengths.

\section{Reinforced dynamics of a D-level (qudit) system}\label{S4}
In this section we consider a qudit where the Hilbert space is spanned by $D$ orthonormal states $\{|d\rangle: d=0,\dots,D-1\}$.
Let us represent the target state by $|\psi_f\rangle=|0\rangle$. In this way $|\psi_f^{\perp}\rangle =\frac{1}{\sqrt{D-1}}\sum_{d=1}^{D-1}|d\rangle$ and $|\psi_i\rangle=\sqrt{P_0}|\psi_f\rangle+\sqrt{1-P_0}|\psi_f^{\perp}\rangle$ with $P_0=1/D$. We also define the generalized Pauli operators
\begin{align}
	X &=\sum_{d=0}^{D-1} |d+1 \rangle\langle d|,\\
	Z &=\sum_{d=0}^{D-1} \omega^d|d \rangle\langle d|,\\
	Y &=XZ,
\end{align}
where $\omega=e^{\mathrm{i}2\pi/D}$ and $D \equiv 0$ (modulo $D$). To get rid of the statistical errors and reduce the computation time in the following we work with deterministic perturbations.

\subsection{Coherent noise}\label{S41} 
As before the reinforced Hamiltonian is given by $H_l(r)=A_lH_i+B_lH_f-r_l\rho_l+V_l$ with quantum state $\rho_l =|\psi_l\rangle\langle \psi_l|$. We use the shift operator to define the coherent noise
\begin{align}
	V_l = \epsilon_l (X+X^{\dagger}),
\end{align}
with $\epsilon_l=\epsilon/L$. Here $r_l=r$ for all layers and $|\psi_{l+1}\rangle=U_l(r)|\psi_{l}\rangle$ starting from $|\psi_i\rangle$.

\begin{figure}
	\includegraphics[width=16cm]{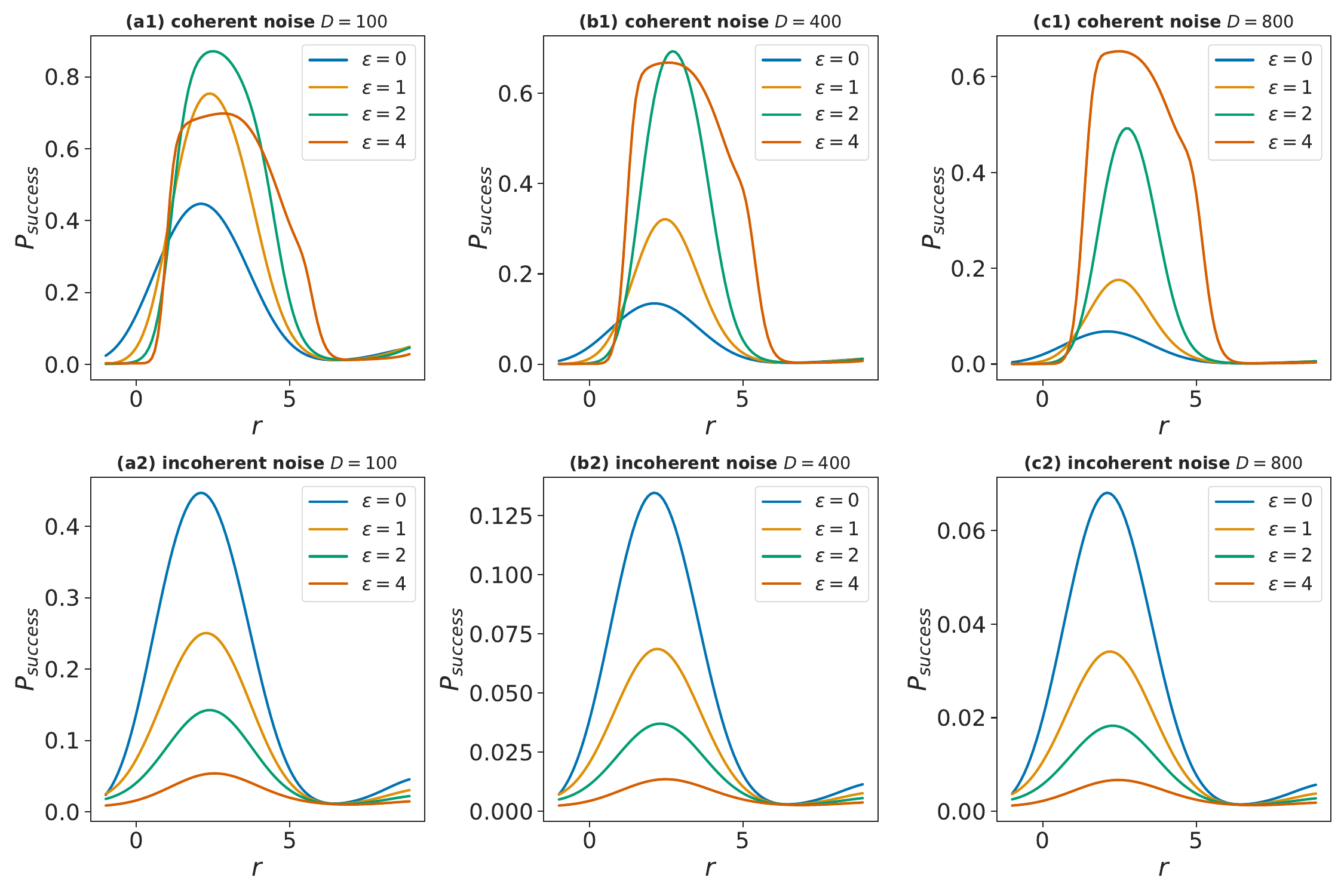} 
	\caption{Quantum annealing with coherent and incoherent noise in a $D$-level (qudit) system using the Grover coefficients $A_l^G,B_l^G$: Success probability $P_{success}(L)$ vs reinforcement parameter $r$ for different noise strengths $\epsilon$. ((a1),(a2)) $D=100$, ((b1),(b2)) $D=400$, and ((c1),(c2)) $D=800$ for a fixed number of layers $L=10$.}\label{fig3}
\end{figure}

Figure \ref{fig3} (upper panels) displays the success probability $P_{success}(L)$ vs the reinforcement parameter $r$ in a quantum annealing process using the Grover coefficients. As before we set $r_l=r$ for all layers. 
The figure shows the results for different dimensions $D=100,400,800$ and values of noise strength $\epsilon=0,1,2,4$, in order to identify the optimal reinforcement strength. Note that the optimal reinforcement is $r\simeq 2.5$ and it is independent of the reported parameters $D$ and $\epsilon$. The good news is that we do not have to increase the strength of reinforcement for higher dimensions and noise strengths.   

In Fig. \ref{fig3}, we observe a constructive effect between the reinforcement and the perturbation 
introduced by the $X$ operator: the optimal success probability initially rises with the strength of $V_l$. This 
behavior does not occur for other perturbations involving the $Y$ and $Z$ operators (not shown here). The goal of 
quantum annealing is to follow the instantaneous ground state of $H_l^{QA}=A_lH_i+B_lH_f$. Note that $V_l$, as a shift operator, corresponds to a symmetric random walk that tends to spread the state across the dimension space $d=0,\dots,D-1$. In contrast, the reinforcement term attempts to project the dynamics onto the current state of the system. Consequently, the reinforcement term together with the random-walk contribution $V_l-r\rho_l$ induce a biased walk that favors the ground state of $H_l^{QA}$. This behavior can improve the success probability when $r$ and $\epsilon$ are appropriately chosen. A large value of $r$ results in a strong projection onto previous states, thereby missing the annealing dynamics. High values of $\epsilon$, on the other hand, produce a strong diffusion that can drive the system away from the ground-state subspace. This suggests that controlled coherent perturbations may be useful when combined with reinforcement, a point that deserves further study.

\begin{figure}
	\includegraphics[width=14cm]{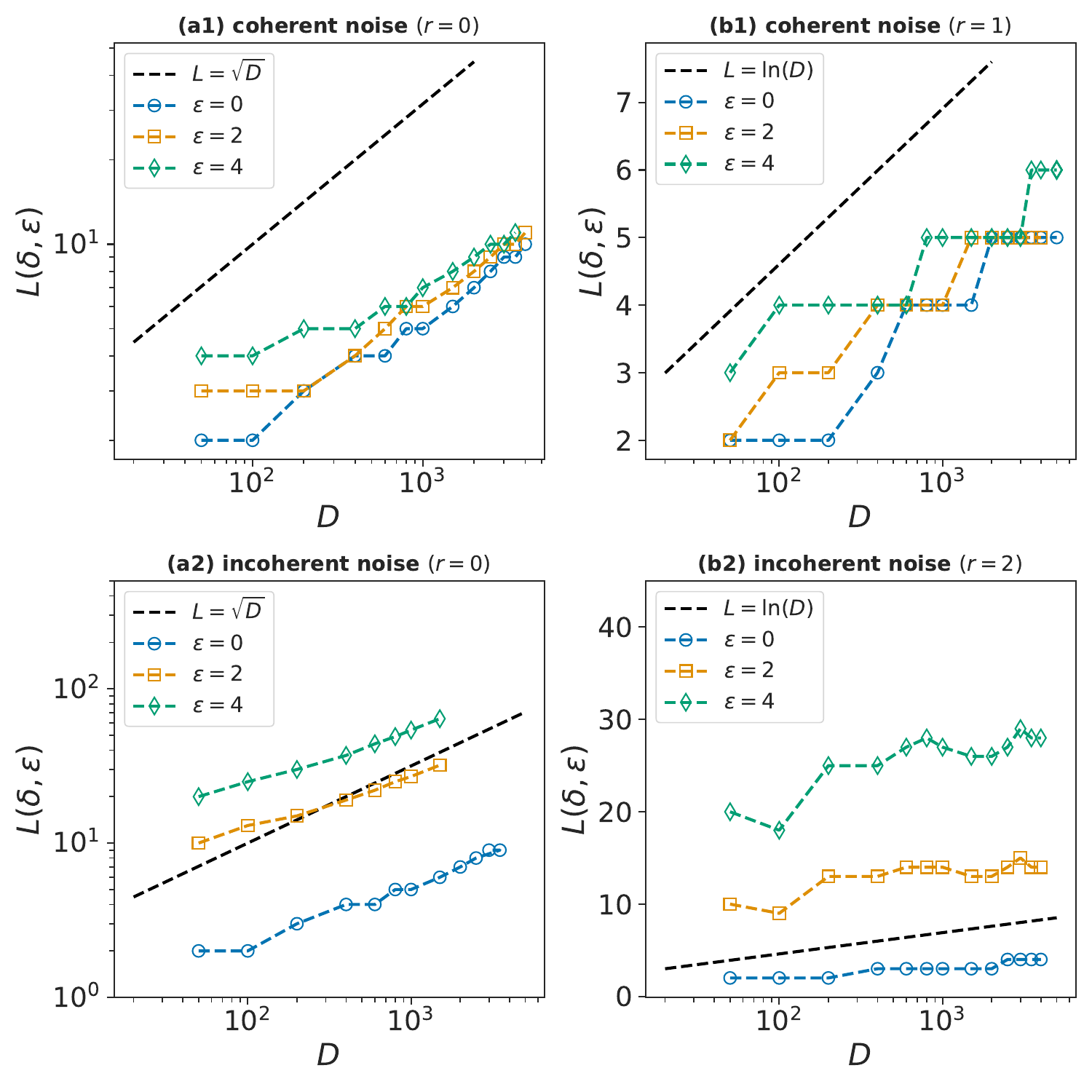} 
	\caption{Quantum search with coherent and incoherent noise in a $D$-level (qudit) system using the locally optimal coefficients $A_l^*,B_l^*$. The computation time $L(\delta,\epsilon)$ is plotted vs dimension $D$ for different noise strengths $\epsilon$. The $L(\delta,\epsilon)$ is the minimum number of evolution layers needed to have $P_{success}>1-\delta$. Here $\delta=1/3$. Panels (a1) and (b1) show the results for coherent noise without reinforcement $r=0$ and with reinforcement $r=1$, respectively. Panels (a2) and (b2) show the results for incoherent noise without reinforcement $r=0$ and with reinforcement $r=2$, respectively.}\label{fig4}
\end{figure}

To assess the scaling of the required number of layers, we obtain the success probability of the quantum dynamics using locally optimal coefficients $A_l^*, B_l^*$. At each layer, we perform an exhaustive search over $K^2$ discrete values of $(A_l,B_l)$ within the range $(0,2\pi)^2$. Here we take $K=10$. Figure \ref{fig4} (upper panels) shows the minimum number of reinforced evolution layers needed to have a success probability that is greater than $1-\delta$ in the process. The results for $\epsilon=0,2,4$ and dimensions $D\in (50,5000)$ support the reduction of the number of layers with reinforcement in the presence of coherent perturbations. Even though the dimension is restricted to small values, we observe convergence toward the expected scaling relations from the noise-free dynamics ($\epsilon = 0$). To better show the scaling behavior of the data, we use log-log plots for $r=0$ and linear-log plots for $r=1$.

\subsection{Incoherent noise}\label{S42} 
The incoherent noise here is modeled by the shift operator $X$ acting on the system alongside the unitary evolution. More precisely, the quantum state evolves according to
\begin{align}
	\rho_{l+1} =(1-\epsilon_l)U_l\rho_l U_l^{\dagger}+\epsilon_l X(U_{l}\rho_l U_{l}^{\dagger})X^{\dagger},
\end{align}
where $U_l$ represents a unitary evolution with reinforced Hamiltonian $H_l =A_lH_i+B_lH_f-r\rho_l$. Again we set $\epsilon_l=\epsilon/L$ and $r_l=r$ for all layers. The success probability at any layer $l$ is given by $P_{success}(l)=\mathrm{tr}(\rho_{l}|0 \rangle\langle 0|)$. 

Figure \ref{fig3} (lower panels) shows the success probability vs the reinforcement parameter $r$ at the end of a quantum annealing process using the Grover coefficients. Like the coherent case we observe an optimal $r\simeq 2.5$ that it is not sensitive to values of dimension $D$ and noise strength $\epsilon$. Next we consider the reinforced dynamics with the locally optimal coefficients $(A_l^*, B_l^*)$ to study the scaling of the required number of layers with the system dimension. Here we set the reinforcement parameter to $r=2$, as opposed to $r=1$ for the coherent case, because incoherent noise is more difficult to mitigate with reinforcement. Figure \ref{fig4} (lower panels) shows the minimum $L$ for which $P_{success}(l)$ exceeds $1-\delta$ at some layer $l$ during the process. The required numbers of evolution layers are higher than those obtained in the presence of coherent noise; however, we again observe a substantial reduction of the number of layers due to reinforcement.

\section{Conclusion}\label{S5}
We have numerically shown that, in the absence of noise, reinforcement reduces the number of layers required for the quantum search dynamics by using an effectively nonlinear, state-dependent evolution which has access to the system state. As a result, the success probability of a quantum search algorithm is significantly enhanced by reinforcement in the presence of coherent and incoherent noise. The study of noisy quantum evolutions was limited to small systems of $N\le 12$ qubits and $D\le 5000$ for the $D$-level (qudit) system \cite{github-link}.
It would be interesting to see how an approximate treatment of such reinforced evolutions in larger systems affects the performance of a quantum search or optimization algorithm. As a byproduct, we also observed that combining reinforcement with a coherent perturbation implemented via the shift operator $X$ can constructively increase the success probability of the annealing algorithm. This effect should be explored independently in future work to assess its computational benefits for optimization using quantum annealing algorithms.

We emphasize that the reported reduction in the number of evolution layers does not contradict Grover optimality, which applies to standard linear-oracle/query models of unstructured search. The reinforced evolution is state-dependent and assumes access to additional information, namely an estimate of the instantaneous system state. The results should therefore be understood as a dynamical speedup within this feedback-assisted model.

Finding an efficient and approximate representation of reinforcement is essential for implementing a reinforced quantum algorithm. The number of system copies that are needed for an exact quantum state tomography grows exponentially with the number of qubits \cite{Haah-acm-2016}. However, an approximate estimation of the quantum state from the expectation values of a polynomial number of observables would be less expensive; in a collective weak measurement, $M\propto 1/\delta$ identical copies of the system are needed to estimate the expectation value of an observable with accuracy $\epsilon\propto 1/\sqrt{M}$, while perturbing the quantum state of a single system by $\delta$ \cite{lloyd-pra-2000}. This implies that approximating $\rho_l$ using only one-qubit Pauli operators requires roughly $M\propto (3N)L/\epsilon^2$ copies for $L$ layers of evolution in an $N$-qubit system. This mean-field approximation of $\rho_l$ can still perform well in some optimization problems, as demonstrated in Ref. \cite{QMC-pra-2018}.     

In practice, one may employ a Trotter decomposition of the unitary $U_l(r)\approx (e^{-\mathrm{i}H_l(r)/M})^M$ in terms of local unitary operators. The main difficulty clearly lies with the reinforcement term $e^{\mathrm{i}(r/M)\rho_l}$. It is known that $M$ copies of the quantum state $\rho_l$ are sufficient to implement the unitary $e^{-\mathrm{i}\rho_l M\Delta t}$ with an error of order $M\Delta t^2$ \cite{lloyd-nphys-2014}. Here $|\Delta t|=r/M$ and the error scales as $\epsilon=r^2/M$. Consequently, the number of copies required for $L$ layers is $M\propto r^2L/\epsilon$, which depends on the scaling of $L$ with the system size $N$.  

Another strategy is to model the nonlinear reinforced dynamics $U_l(r)=e^{-\mathrm{i}H_l(r)}$ by an effective linear evolution $\tilde{U}_l=e^{-\mathrm{i}\tilde{H}_l}$. The parameters of the local Hamiltonian $\tilde{H}_l$ can be inferred via a learning algorithm, as described in Ref. \cite{QR-jstat-2025}. The computational cost of this inference for local and sparse Hamiltonians $\tilde{H}_l$ scales as $L^2N^2$. Note, however, that all the above approaches rely on an approximation of the reinforced quantum evolution, which may compromise the performance of the original  algorithm. This underscores the importance of developing accurate approximate methods for handling such reinforced quantum dynamics.

\acknowledgments
We would like to thank Mohammad Hossein Zarei for helpful discussions.
This work was performed using the ALICE computing resources provided by Leiden University.

\end{document}